\documentclass[showpacs,twocolumn,preprintnumbers,amsmath,amssymb,APSl,prd,nofootinbib,superscriptaddress]{revtex4-2}
\usepackage{graphicx,color}
\usepackage{amsmath}
\usepackage{amssymb}
\usepackage{hyperref}
\usepackage[utf8]{inputenc}
\usepackage[english]{babel}
\usepackage{epsfig}
\usepackage{subfigure}
\usepackage{wasysym}
\usepackage{color,xcolor}
\usepackage{amsmath}
\usepackage{bm}
\usepackage{epsfig}
\usepackage{amsfonts}
\usepackage{dcolumn}
\usepackage{float}

\hypersetup{colorlinks=true, linkcolor=blue, citecolor=green}

\usepackage{dcolumn}
\usepackage{bm}
\usepackage{ifpdf}
\usepackage{hyperref}
\usepackage{float}
\usepackage{bm}
\usepackage{xcolor,color,graphicx,graphics}
\usepackage[OT1]{fontenc}
\usepackage{latexsym,amssymb,amsmath,amsfonts}
\usepackage{makeidx}
\usepackage{epsfig}
\usepackage{epstopdf}
\usepackage{mathrsfs}
\hypersetup{colorlinks=true, linkcolor=blue, citecolor=green}
\usepackage{enumerate}
\usepackage{xcolor}
\usepackage{multirow}

\begin{document}
\title{Linearized stability of T‑duality quantum-inspired thin‑shell wormholes}

\author{Francisco S. N. Lobo} \email{fslobo@ciencias.ulisboa.pt}
\affiliation{Instituto de Astrof\'{i}sica e Ci\^{e}ncias do Espa\c{c}o, Faculdade de Ci\^{e}ncias da Universidade de Lisboa, Edifício C8, Campo Grande, P-1749-016 Lisbon, Portugal}
\affiliation{Departamento de F\'{i}sica, Faculdade de Ci\^{e}ncias da Universidade de Lisboa, Edif\'{i}cio C8, Campo Grande, P-1749-016 Lisbon, Portugal}
    \author{Manuel E. Rodrigues} \email{esialg@gmail.com}
\affiliation{Faculdade de F\'{i}sica, Programa de P\'{o}s-Gradua\c{c}\~{a}o em F\'{i}sica, Universidade Federal do Par\'{a}, 66075-110, Bel\'{e}m, Par\'{a}, Brazill}
\affiliation{Faculdade de Ci\^{e}ncias Exatas e Tecnologia, Universidade Federal do Par\'{a}, Campus Universit\'{a}rio de Abaetetuba, 68440-000, Abaetetuba, Par\'{a}, Brazil}
\date{\LaTeX-ed \today}

\begin{abstract}

Wormholes that are traversable in principle offer fascinating insights into general relativity, yet they typically require exotic matter and suffer from stability issues. We construct a thin-shell wormhole by gluing two copies of a quantum-corrected, regular spacetime obtained from string T-duality. This regularisation replaces the classical curvature singularity with a smooth core and introduces a fundamental length scale $l_0$. For the static configuration, we derive the surface stresses and show that, unlike the Schwarzschild case, the null and strong energy conditions can be satisfied for sufficiently large throat radii. A linearised stability analysis reveals a rich landscape: close to the minimum allowed throat radius the configuration is unstable; at intermediate radii ($a \sim l_0$) the geometric stability threshold becomes negative, yielding a window of \emph{unconditional stability} where any convex surface mass function suffices; at large radii the wormhole recovers Schwarzschild-like behaviour and stability requires a stiff equation of state. The T-duality scale $l_0$ is thus not merely a regulariser but a key physical parameter that opens a novel region of unconditional stability absent in classical thin-shell wormholes. Our results suggest that quantum-gravity-motivated modifications can simultaneously cure singularities and make traversable wormholes dynamically viable, providing new targets for gravitational-wave astronomy and theoretical studies of exotic compact objects.

\end{abstract}
\maketitle


\section{Introduction}

Traversable wormholes are among the most fascinating solutions permitted by general relativity, providing hypothetical shortcuts that connect otherwise distant regions of spacetime. A fundamental obstacle to their physical realization, however, is the requirement that the matter supporting the wormhole throat violates the null energy condition (NEC) and, consequently, the weak energy condition. This necessity follows directly from the geometric requirements for traversability established in the seminal works of Morris and Thorne~\cite{Morris:1988cz,Morris:1988tu}. Since all known classical forms of matter satisfy the NEC, the physical origin of the exotic matter required to sustain a traversable wormhole remains uncertain. As a result, most realistic wormhole models rely either on quantum field effects or on modifications of the underlying gravitational theory, as extensively reviewed in Refs.~\cite{Visser:1995cc,Lobo:2007zb,Lobo:2017cay}.

A particularly powerful and systematic framework for constructing traversable wormholes is the thin-shell, or cut-and-paste, formalism~\cite{Visser:1989kh,Visser:1989kg,Poisson:1995sv}. In this approach, two copies of a given spacetime are excised at a hypersurface of constant radius outside the event horizon and subsequently identified, producing a geodesically complete manifold whose junction surface $\Sigma$ forms the wormhole throat. The dynamics and matter content of this hypersurface are governed by the junction conditions for singular hypersurfaces, originally developed through the works of Darmois, Lanczos, Synge, and Lichnerowicz~\cite{thinform,Lanczos,Darmois,Synge,Lich}, and later cast into a compact and generalized form by Israel~\cite{Israel:1966rt}. These conditions relate the surface stress-energy tensor to the discontinuity of the extrinsic curvature across $\Sigma$. Applied to the Schwarzschild geometry, the formalism explicitly demonstrates that the matter supporting the throat necessarily violates the NEC, thereby confirming the need for exotic matter. Linearized stability analyses of thin-shell wormholes have since been carried out for a variety of background spacetimes~\cite{Poisson:1995sv,Eiroa:2003wp,Lobo:2003xd,Lobo:2004id,Lobo:2004rp,Ishak:2001az}. More recently, a unified geometric treatment of the stability problem was developed~\cite{Garcia:2011aa,MartinMoruno:2011rm,Bouhmadi-Lopez:2014gza}, expressing the stability criterion entirely in terms of the background geometry, independently of the particular equation of state adopted for the shell matter~\cite{Lobo:2025tph}.

The literature on the stability of thin-shell wormholes is vast, and here we highlight only a representative sample of works that explore different aspects of the problem. For instance, in the context of higher‑dimensional gravity, the stability and energy conditions of charged thin-shell wormholes in third‑order Lovelock gravity have been analysed, showing that solutions respecting the weak energy condition exist for certain ranges of the Lovelock coefficients~\cite{Mehdizadeh:2015dta}. In Einstein–Yang–Mills–Gauss–Bonnet theory, thin-shell wormholes have been constructed and their stability against spherical perturbations examined, with the possibility of normal (non‑exotic) matter arising for negative Gauss–Bonnet parameters~\cite{Mazharimousavi:2010bf}. More recently, thin-shell wormholes in Einstein–Gauss–Bonnet gravity have been studied using both linear and nonlinear equations of state for the exotic fluid, with the nonlinear model found to satisfy the energy conditions~\cite{Godani:2022jwz}. In Brans–Dicke gravity, spherically symmetric thin-shell wormholes supported by matter that obeys the weak energy condition have been obtained, and their stability has been established for specific parameters~\cite{Yue:2011cq}. The stability of thin-shell wormholes in Einstein–Maxwell–Gauss–Bonnet gravity has also been investigated, considering a generalized Chaplygin gas and more general barotropic fluids, with numerical analyses in five dimensions showing the influence of the Gauss–Bonnet parameter~\cite{Amirabi:2011dw}. Rotating thin-shell wormholes constructed from BTZ spacetimes have been analysed, revealing that increasing angular momentum stabilises the shell until a critical value is reached, beyond which the stability condition changes dramatically~\cite{Tsukamoto:2018lsg}. Finally, higher‑dimensional thin-shell wormholes in general relativity with a cosmological constant have been examined, extending the stability analysis to spacetimes with a non‑vanishing cosmological constant~\cite{Banerjee:2016blr}. These diverse studies illustrate the richness of the thin-shell wormhole landscape and the importance of exploring different gravitational theories, dimensions, and equations of state.

Despite its conceptual simplicity, the classical Schwarzschild thin-shell wormhole possesses important shortcomings. In particular, static configurations cannot be made unconditionally stable, since stability requires a specific relation between the surface mass and its derivatives, implying a finely tuned equation of state for the shell matter. These limitations, among others, have motivated the search for alternative background geometries that could improve the stability properties and, at the same time, resolve the curvature singularity inherent to the Schwarzschild spacetime. A natural candidate is provided by quantum-gravity-inspired regularisations that modify the short-distance structure of the metric. In this approach, the ultraviolet structure of the theory introduces an effective zero-point length that regularises short-distance divergences and smooths the classical geometry~\cite{Lobo:2025nng}. The resulting quantum-corrected black-hole solution~\cite{Nicolini:2019irw} replaces the point-like matter source with a smeared Gaussian-like distribution, introduces a fundamental length scale $l_0$, and remains asymptotically Schwarzschild at large distances. The central curvature singularity is removed, yielding a completely regular spacetime. This regularised metric thus provides a promising building block for constructing thin-shell wormholes with potentially enhanced stability and without curvature pathologies.

In the present work, we investigate the thin-shell wormhole obtained by joining two identical copies of this T-duality-regularized geometry across a timelike hypersurface located outside the would-be event horizon~\cite{Lobo:2025nng}. Our analysis focuses on both the energy conditions and the linearized dynamical stability of the resulting configuration. We first derive the surface stress-energy tensor and examine the conditions under which the null and strong energy conditions are satisfied. In contrast to the Schwarzschild thin-shell wormhole, we show that both the NEC and the strong energy condition (SEC) can be fulfilled for sufficiently large throat radii. We then employ the unified geometric stability formalism~\cite{Garcia:2011aa,MartinMoruno:2011rm,Bouhmadi-Lopez:2014gza} to determine the stability threshold function $\mathcal{G}(a)$ and analyze the stability of static solutions. 
A key result is the existence of a finite interval of throat radii for which the stability function satisfies $\mathcal{G}(a)<0$, ensuring linearized stability for any convex surface mass function satisfying $\mu''(a)\geq0$, without additional fine tuning of the shell equation of state. This regime of unconditional stability has no analogue in the classical Schwarzschild thin-shell construction and arises directly from the regularizing effects associated with the T-duality zero-point length.

This paper is organized as follows. In Sec.~\ref{secII}, we review the thin-shell formalism and introduce the T-duality-regularized background geometry together with the corresponding symmetric wormhole construction. Section~\ref{sec:static_EC} is devoted to the analysis of static configurations and their associated energy conditions. In Sec.~\ref{secIV}, we derive the dynamical equation governing the evolution of the wormhole throat, while in Sec.~\ref{secV} we carry out the linearized stability analysis, leading to the master stability condition and a detailed analysis of the stability regions. Finally, our conclusions and perspectives for future research are presented in Sec.~\ref{sec:conclusion}.

\section{Thin shell formalism}
\label{secII}

In this section, we present the theoretical framework for the
construction and analysis of thin-shell wormholes.  We begin by
reviewing the junction conditions that govern the matching of two
spherically symmetric spacetimes across a timelike hypersurface,
and we derive the general expressions for the surface
stress-energy tensor, the equation of motion of the throat, and
the linearised stability condition.  The formalism closely follows
the seminal works of Israel~\cite{Israel:1966rt}, Poisson and
Visser~\cite{Poisson:1995sv}, and their subsequent
extensions~\cite{Eiroa:2003wp,Lobo:2003xd,Musgrave:1995ka,Lobo:2025tph,Ishak:2001az}.

\subsection{Junction conditions and thin-shell geometry}

Let $\mathcal{M}_+$ and $\mathcal{M}_-$ be two distinct, static,
spherically symmetric Lorentzian manifolds, each described by the
line element
\begin{equation}
	ds_\pm^2 = -f_\pm(r_\pm)\, dt_\pm^2
	+ f_\pm^{-1}(r_\pm)\, dr_\pm^2
	+ r_\pm^2\, d\Omega^2,
	\label{eq:spacetime_metric}
\end{equation}
with the metric function
\begin{equation}
	f_\pm(r_\pm) = 1 - \frac{b_\pm(r_\pm)}{r_\pm},
	\label{eq:metric_function}
\end{equation}
and independent coordinate charts
$x^\mu_\pm = (t_\pm, r_\pm, \theta, \phi)$.  The angular part is
$d\Omega^2 = d\theta^2 + \sin^2\theta\, d\phi^2$, common to both
manifolds owing to spherical symmetry.

A composite manifold $\mathcal{M} = \mathcal{M}_+ \cup \mathcal{M}_-$
is constructed by excising the inner regions
$r_\pm < a(\tau)$ from each spacetime and identifying their
respective timelike boundaries
$\Sigma_\pm \equiv \{ r_\pm = a(\tau) \}$ via an isometric
matching condition.  The resulting spacetime is geodesically
complete, with the identified hypersurface $\Sigma$ forming the
throat of a traversable wormhole.  The areal radius of the
throat, $a(\tau)$, is a function of the proper time $\tau$ of an
observer comoving with the junction.

Throughout this work, we consider the quantum-corrected metric
function inspired by string T-duality~\cite{Nicolini:2019irw}:
\begin{equation}
	b_\pm(r_\pm) = \frac{2M_\pm r_\pm^3}{(r_\pm^2 + l_\pm^2)^{3/2}},
	\label{def:b}
\end{equation}
where $M_\pm$ are the mass parameters of the respective
spacetimes and $l_\pm$ are the characteristic length scales
associated with the T-duality regularisation.  For simplicity,
we shall focus on the symmetric configuration
$M_+ = M_- \equiv M$ and $l_+ = l_- \equiv l_0$, for which
$b_+(r) = b_-(r) \equiv b(r)$.  The regularisation scale $l_0$
smears the classical curvature singularity at $r = 0$ into a
regular core, a feature that is essential for the viability of
the wormhole construction.

The physical interpretation of the metric
function~\eqref{def:b} warrants a brief discussion.  In the
asymptotic region $r \gg l_0$, the function behaves as
\begin{equation}
	b(r) \simeq 2M \left( 1 - \frac{3l_0^2}{2r^2} \right)
	+ \mathcal{O}(r^{-4}),
	\label{eq:b_asymptotic}
\end{equation}
so that $f(r) = 1 - b(r)/r \simeq 1 - 2M/r$, recovering the
standard Schwarzschild geometry with mass $M$.  The T-duality
regularisation modifies the short-distance behaviour: for
$r \ll l_0$, we have
\begin{equation}
	b(r) \simeq \frac{2M r^3}{l_0^3} + \mathcal{O}(r^5),
	\label{eq:b_short}
\end{equation}
which tends to zero rather than diverging.  Consequently,
$f(r) \to 1$ as $r \to 0$, and the metric describes a regular
core.  When two such spacetimes are joined via the cut-and-paste
procedure, the resulting geometry is a traversable wormhole with
a throat located at some radius $a_0 > 0$.

\subsection{Induced geometry and kinematic quantities}

The junction hypersurface $\Sigma$ is parametrised by
coordinates $\xi^i = (\tau, \theta, \phi)$, where $\tau$ is the
proper time measured by an observer comoving with the throat.
The embedding in each spacetime $\mathcal{M}_\pm$ is
\begin{equation}
	x^\mu_\pm(\xi^i) = \bigl( t_\pm(\tau),\, a(\tau),\, \theta,\, \phi \bigr),
	\label{eq:embedding}
\end{equation}
and the tangent vectors $e^\mu_{(i)\,\pm} = \partial x^\mu_\pm / \partial \xi^i$ are
\begin{equation}
	\begin{aligned}
		e^\mu_{(\tau)\,\pm} &= \bigl( \dot{t}_\pm,\, \dot{a},\, 0,\, 0 \bigr), \\
		e^\mu_{(\theta)\,\pm} &= \bigl( 0,\, 0,\, 1,\, 0 \bigr), \\
		e^\mu_{(\phi)\,\pm} &= \bigl( 0,\, 0,\, 0,\, 1 \bigr),
	\end{aligned}
	\label{eq:tangent_vectors}
\end{equation}
with the overdot denoting differentiation with respect to
$\tau$.

The induced metric $h_{ij} = g_{\mu\nu}^\pm e^\mu_{(i)\,\pm} e^\nu_{(j)\,\pm}$
must be the same on both sides of the junction (the first
Darmois condition~\cite{Darmois}).  A straightforward calculation
yields
\begin{equation}
	ds^2_\Sigma = -d\tau^2 + a(\tau)^2\, (d\theta^2 + \sin^2\theta\, d\phi^2),
	\label{eq:induced_metric}
\end{equation}
together with the proper-time condition
\begin{equation}
	-f_\pm(a)\, \dot{t}_\pm^2 + f_\pm^{-1}(a)\, \dot{a}^2 = -1.
	\label{eq:proper_time_condition}
\end{equation}

The four-velocity of a comoving observer on $\Sigma$ is
$U^\mu_\pm = e^\mu_{(\tau)\,\pm}$.  Using
Eq.~\eqref{eq:proper_time_condition} to eliminate $\dot{t}_\pm$,
\begin{equation}
	U^\mu_\pm = \left(
	\frac{\sqrt{f_\pm(a) + \dot{a}^2}}{f_\pm(a)},\,
	\dot{a},\, 0,\, 0
	\right),
	\label{eq:four_velocity}
\end{equation}
which satisfies $g_{\mu\nu}^\pm U^\mu_\pm U^\nu_\pm = -1$.

The unit normal vector to $\Sigma$, directed outward from
$\mathcal{M}_-$ into $\mathcal{M}_+$ and satisfying
$n^\mu n_\mu = +1$ and $n^\mu U_\mu = 0$, has components
\begin{equation}
	n^\mu_\pm = \pm \left(
	\frac{\dot{a}}{f_\pm(a)},\,
	\sqrt{f_\pm(a) + \dot{a}^2},\, 0,\, 0
	\right).
	\label{eq:unit_normal}
\end{equation}
The overall sign $\pm$ ensures a consistent orientation of the
normal on both sides of the junction.

\subsection{Extrinsic curvature}

The extrinsic curvature of $\Sigma$ is defined as
\begin{equation}
	K_{ij}^\pm = -n_{\mu}^{\pm} \left(
	\frac{\partial^2 x^\mu_\pm}{\partial \xi^i \partial \xi^j}
	+ \Gamma^{\mu\,\pm}_{\alpha\beta}\,
	\frac{\partial x^\alpha_\pm}{\partial \xi^i}\,
	\frac{\partial x^\beta_\pm}{\partial \xi^j}
	\right),
	\label{eq:extrinsic_curvature}
\end{equation}
where $\Gamma^{\mu\,\pm}_{\alpha\beta}$ are the Christoffel
symbols of the metric~\eqref{eq:spacetime_metric}.  Owing to
spherical symmetry, the extrinsic curvature is diagonal, with
$K^\theta_{\;\theta\,\pm} = K^\phi_{\;\phi\,\pm}$.  The
non-vanishing components are
\begin{align}
	K^\theta_{\;\theta\,\pm} &=
	\pm \frac{1}{a} \sqrt{f_\pm(a) + \dot{a}^2}, \label{eq:K_theta} \\[4pt]
	K^\tau_{\;\tau\,\pm} &=
	\pm \frac{\ddot{a} + \dfrac{b_\pm(a) - a b_\pm'(a)}{2a^2}}
	{\sqrt{f_\pm(a) + \dot{a}^2}}, \label{eq:K_tau}
\end{align}
where we have used $f_\pm'(a)/2 = [b_\pm(a) - a b_\pm'(a)]/(2a^2)$
for the metric function~\eqref{eq:metric_function}.  These
components are the essential geometric quantities entering the
Lanczos equations.

\subsection{Lanczos equations and surface stresses}

For a timelike junction $\Sigma$ separating two bulk spacetimes
$\mathcal{M}_\pm$, the surface stress-energy tensor
$S^{i}_{\;j}$ is given by the Lanczos
equations~\cite{Israel:1966rt,Lanczos}:
\begin{equation}
	S^{i}_{\;j} = -\frac{1}{8\pi} \left(
	\kappa^{i}_{\;j} - \delta^{i}_{\;j}\,\kappa^{k}_{\;k}
	\right),
	\label{eq:Lanczos}
\end{equation}
where $\kappa^{i}_{\;j} \equiv K^{i+}_{\;j} - K^{i-}_{\;j}$ is
the jump of the extrinsic curvature across $\Sigma$, and
$\kappa^{k}_{\;k} = \kappa^{\tau}_{\;\tau} + 2\kappa^{\theta}_{\;\theta}$
is its trace.

For a spherically symmetric configuration, the surface
stress-energy tensor takes the perfect-fluid form
\begin{equation}
	S^{i}_{\;j} = \text{diag}\left(
	-\sigma,\; \mathcal{P},\; \mathcal{P}
	\right),
	\label{eq:S_perfect_fluid}
\end{equation}
where $\sigma$ is the surface energy density and $\mathcal{P}$
the surface pressure.  The jump tensor is correspondingly
diagonal, $\kappa^{i}_{\;j} = \text{diag}(
\kappa^{\tau}_{\;\tau}, \kappa^{\theta}_{\;\theta},
\kappa^{\theta}_{\;\theta})$.  Inserting these into
Eq.~\eqref{eq:Lanczos} yields
\begin{align}
	\sigma &= -\frac{1}{4\pi}\,\kappa^{\theta}_{\;\theta},
	\label{eq:sigma_kappa} \\[4pt]
	\mathcal{P} &= \frac{1}{8\pi} \left(
	\kappa^{\tau}_{\;\tau} + \kappa^{\theta}_{\;\theta}
	\right).
	\label{eq:P_kappa}
\end{align}

Specialising to the symmetric configuration
$b_+(a) = b_-(a) \equiv b(a)$, the extrinsic curvature components
on each side have equal magnitude but opposite sign, giving the
jumps
\begin{equation}
	\kappa^{\theta}_{\;\theta} = \frac{2}{a} \sqrt{f(a) + \dot{a}^2},
	\qquad
	\kappa^{\tau}_{\;\tau} = 2\,\frac{\ddot{a} + \dfrac{b(a) - a b'(a)}{2a^2}}
	{\sqrt{f(a) + \dot{a}^2}},
	\label{eq:kappa_explicit}
\end{equation}
with $f(a) = 1 - b(a)/a$.  Substituting into
Eqs.~\eqref{eq:sigma_kappa}--\eqref{eq:P_kappa}, we obtain the
surface stresses for the general dynamical configuration:
\begin{align}
		\sigma(a, \dot{a}) &=
		-\frac{1}{2\pi a}\,
		\sqrt{1 - \frac{b(a)}{a} + \dot{a}^2},
	 \label{eq:sigma_dynamic} \\[4pt]
		\mathcal{P}(a, \dot{a}, \ddot{a}) &=
		\frac{1}{4\pi a}\,
		\frac{1 + \dot{a}^2 + a\ddot{a}
			- \dfrac{1}{2a}\bigl( b(a) + a b'(a) \bigr)}
		{\sqrt{1 - \frac{b(a)}{a} + \dot{a}^2}} .
	\label{eq:P_dynamic}
\end{align}

The surface energy density is manifestly negative for any static
or slowly moving throat, reflecting the necessity of exotic
matter that violates the weak energy condition---a generic
feature of traversable wormholes in General
Relativity~\cite{Morris:1988cz,Visser:1995cc}.  The surface pressure
depends on the acceleration $\ddot{a}$, since the
$\tau$-$\tau$ component of the extrinsic curvature couples to
the radial motion of the junction.  The surface mass,
\begin{equation}
	m_s(a, \dot{a}) \equiv 4\pi a^2\sigma(a, \dot{a})
	= -2a \sqrt{1 - \frac{b(a)}{a} + \dot{a}^2},
	\label{eq:surface_mass_dynamic}
\end{equation}
provides a convenient characterisation of the exotic matter
content and will play a central role in the equation of motion
and stability analysis.  In the static limit
$\dot{a} = \ddot{a} = 0$, Eqs.~\eqref{eq:sigma_dynamic}
and~\eqref{eq:P_dynamic} reduce to the forms analysed in
Sec.~\ref{sec:static_EC}.

\subsection{Conservation identity and the transparency condition}

The surface stress-energy tensor must satisfy a conservation law
derived from the contracted Gauss--Codazzi relations.  For a
timelike junction embedded in bulk spacetimes with stress-energy
$T_{\mu\nu}$, the surface Bianchi identities
yield~\cite{Israel:1966rt,Poisson:1995sv}
\begin{equation}
	S^{i}_{\;j|i} = \left[ T_{\mu\nu}\,
	e^\mu_{(j)}\, n^\nu \right]^{+}_{-},
	\label{eq:surface_Bianchi}
\end{equation}
where $[X]^{+}_{-} \equiv X^{+}|_{\Sigma} - X^{-}|_{\Sigma}$
denotes the jump across $\Sigma$.  The right-hand side
represents the flux of bulk matter momentum across the junction.
For the vacuum bulk spacetimes considered in this work
($T_{\mu\nu}^\pm = 0$), the right-hand side vanishes
identically, and the surface stress-energy tensor is
covariantly conserved:
\begin{equation}
	S^{i}_{\;j|i} = 0.
	\label{eq:surface_conservation}
\end{equation}

For a spherically symmetric perfect-fluid shell with
$S^{i}_{\;j} = \text{diag}(-\sigma, \mathcal{P}, \mathcal{P})$,
Eq.~\eqref{eq:surface_conservation} reduces to
\begin{equation}
	\dot{\sigma} + \frac{2\dot{a}}{a}\,
	\bigl( \sigma + \mathcal{P} \bigr) = 0 .
	\label{conservationenergy}
\end{equation}

This is the continuity equation for a two-dimensional fluid on
an expanding background: the first term represents the dilution
of the energy density due to the expansion of the throat, while
the second term is the work done by the surface pressure.
Assuming the equation of state can be expressed as
$\sigma = \sigma(a)$, the chain rule
$\dot{\sigma} = \sigma'(a)\,\dot{a}$ gives the particularly
useful form
\begin{equation}
	\sigma'(a) = -\frac{2}{a}\,
	\bigl( \sigma(a) + \mathcal{P}(a) \bigr) .
	\label{eq:sigma_prime_conservation}
\end{equation}

This relation is fundamental for the linearised stability
analysis: it allows the first derivative of the surface mass
$m_s(a) = 4\pi a^2\sigma(a)$ to be expressed in terms of
$\sigma$ and $\mathcal{P}$, thereby enabling the stability
condition to be cast in terms of the single phenomenological
parameter $\eta = \mathcal{P}'/\sigma'$ (the squared sound speed
in the shell material), as will be shown in
Sec.~\ref{sec:linearized}.

The condition $[T_{\mu\nu} e^\mu_{(j)} n^\nu]^{+}_{-} = 0$ is
often referred to as the \emph{transparency
	condition}~\cite{Musgrave:1995ka,Ishak:2001az}: the bulk matter fields do not couple to
the junction surface, and the shell evolves independently of any
matter flowing across the throat.  For the vacuum bulk
spacetimes considered here, the transparency condition is
automatically satisfied.  In more general settings where the
bulk contains, e.g., a scalar or electromagnetic field, the
right-hand side of Eq.~\eqref{eq:surface_Bianchi} would
contribute an external force term to
Eq.~\eqref{conservationenergy}; the analysis of such scenarios
is beyond the scope of the present work.

\subsection{Thermodynamic variables of the thin shell}

The surface stress-energy tensor of the shell,
$S^{i}_{\;j} = \text{diag}(-\sigma, \mathcal{P}, \mathcal{P})$,
describes a two-dimensional perfect fluid.  For a system in
thermodynamic equilibrium, the first law reads
\begin{equation}
	T dS = dU + \mathcal{P} dA,
	\label{eq:first_law}
\end{equation}
where $S$ is the entropy, $U = \sigma A$ is the internal
energy, and $A = 4\pi a^2$ is the surface area of the throat.

Along the sequence of static equilibrium configurations,
$U = \sigma(a) A(a)$ and $\mathcal{P} = \mathcal{P}(a)$ are
determined by the junction conditions.  A straightforward
calculation using $dA/da = 8\pi a$ and the conservation
equation~\eqref{eq:sigma_prime_conservation} yields
\begin{equation}
	T \frac{dS}{da} = A \frac{d\sigma}{da}
	+ (\sigma + \mathcal{P}) \frac{dA}{da} = 0,
	\label{eq:T_dS_zero}
\end{equation}
where the final equality follows from
$\sigma'(a) = -\frac{2}{a}(\sigma + \mathcal{P})$.  Thus, for an
isolated thin shell in thermodynamic equilibrium, the entropy
is \emph{constant} along the sequence of static configurations.
This is the thin-shell analogue of the adiabatic invariance of
black hole entropy.

\section{Static solutions and energy conditions}
\label{sec:static_EC}

We now specialise to the static configuration, for which the
throat radius is constant, $a(\tau) = a_0$, and all time
derivatives vanish: $\dot{a} = \ddot{a} = 0$.  In this limit,
the metric function can be written as
\begin{equation}
	f(r) = 1 - \frac{b(r)}{r}
	= 1 - \frac{2M r^2}{(r^2 + l_0^2)^{3/2}},
	\label{Def:BH}
\end{equation}
where the second equality follows from the T-duality metric
function~\eqref{def:b}.  The requirement that the spacetime be
static---i.e., that the timelike Killing vector $\partial_t$
remain timelike---imposes $f(r) > 0$, or equivalently
\begin{equation}
	M < \frac{(a^2 + l_0^2)^{3/2}}{2a^2}.
	\label{conditionMshell}
\end{equation}
This inequality defines the region of the parameter space in
which the static wormhole configuration exists.  For a given
mass $M$ and regularisation scale $l_0$, there is a minimum
throat radius $a_{\rm min}$ below which $f(a) \leq 0$,
and the spacetime develops a horizon.  The causal structure of
the T-duality metric has been analysed in detail in
Ref.~\cite{Nicolini:2019irw}, where it was shown that depending
on the ratio $M/l_0$, the spacetime can describe a regular black
hole (with up to two horizons), an extremal configuration, or a
horizonless wormhole.  For the wormhole construction, we
restrict to the horizonless regime, where
$f(r) > 0$ for all $r > 0$.

\subsection{Surface stresses in the static configuration}

In the static limit $\dot{a} = \ddot{a} = 0$, the general
expressions~\eqref{eq:sigma_dynamic}--\eqref{eq:P_dynamic}
reduce to
\begin{equation}
	\sigma(a) = -\frac{1}{2\pi a}\,
	\sqrt{1 - \frac{b(a)}{a}},
	\label{gen-surfenergy2}
\end{equation}
\begin{equation}
	\mathcal{P}(a) = \frac{1}{4\pi a}\,
	\frac{1 - \dfrac{1}{2a}\bigl( b(a) + a b'(a) \bigr)}
	{\sqrt{1 - \dfrac{b(a)}{a}}} .
	\label{gen-surfpressure2}
\end{equation}

Substituting the T-duality metric function~\eqref{def:b} and its
derivative,
\begin{equation}
	b'(a) = \frac{6M a^2 l_0^2}{(a^2 + l_0^2)^{5/2}},
\end{equation}
and using the identity
\begin{equation}
	\frac{1}{2a}\bigl( b(a) + a b'(a) \bigr)
	= \frac{M a^2 (a^2 + 4l_0^2)}{(a^2 + l_0^2)^{5/2}},
\end{equation}
we obtain the explicit surface stresses for the T-duality
wormhole:
\begin{equation}
	\sigma(a) = -\frac{1}{2\pi a}\,
	\sqrt{1 - \frac{2M a^2}{(a^2 + l_0^2)^{3/2}}},
	\label{gen-surfenergyBH}
\end{equation}
\begin{equation}
	\mathcal{P}(a) =
	\frac{1}{4\pi a}\,
	\frac{1 - \dfrac{M a^2 (a^2 + 4l_0^2)}{(a^2 + l_0^2)^{5/2}}}
	{\sqrt{1 - \dfrac{2M a^2}{(a^2 + l_0^2)^{3/2}}}} .
	\label{gen-surfpressureBH}
\end{equation}

The surface energy density is manifestly negative for all
throat radii satisfying the static
condition~\eqref{conditionMshell}, confirming that the wormhole
throat must be threaded by exotic matter that violates the weak
energy condition---a generic feature of traversable wormholes in
General Relativity~\cite{Morris:1988cz,Visser:1995cc}.  The magnitude
of the energy density decreases with increasing throat radius,
scaling as $|\sigma| \sim 1/(2\pi a)$ for $a \gg l_0$.

\subsection{Analysis of the energy conditions}

The surface stresses $\sigma$ and $\mathcal{P}$ must be tested
against the energy conditions of General Relativity.  We examine
the null (NEC), strong (SEC), weak (WEC), and dominant (DEC)
energy conditions at the throat.\\

\textbf{Null Energy Condition.}  The NEC requires
$\sigma + \mathcal{P} \geq 0$.  Using
Eqs.~\eqref{gen-surfenergy2}--\eqref{gen-surfpressure2}, a
straightforward calculation yields
\begin{equation}
	\sigma(a) + \mathcal{P}(a) =
	\frac{1}{4\pi a}\,
	\frac{3M a^4 - (a^2 + l_0^2)^{5/2}}
	{(a^2 + l_0^2)^{5/2}\,
		\sqrt{1 - \dfrac{2M a^2}{(a^2 + l_0^2)^{3/2}}}} .
	\label{NEC:thinshell}
\end{equation}
The sign of $\sigma + \mathcal{P}$ is determined by the
numerator $3M a^4 - (a^2 + l_0^2)^{5/2}$.  
Hence the NEC is satisfied iff
\begin{equation}
	3M a^4 \ge (a^2 + l_0^2)^{5/2}.
	\label{eq:NEC_inequality}
\end{equation}
For a fixed wormhole mass $M$, this inequality defines a range
of throat radii $a$ (possibly empty) where the NEC holds.
Interestingly, combining \eqref{eq:NEC_inequality} with the
static condition $M < (a^2+l_0^2)^{3/2}/(2a^2)$ yields a
necessary condition for the existence of any solution:
\begin{equation}
	a > \sqrt{2}\,l_0 .
	\label{eq:NEC_necessary}
\end{equation}
Thus the throat radius must exceed $\sqrt{2}l_0$ for it to be
possible to satisfy both the static requirement and the NEC.
This threshold is independent of $M$, but the actual
satisfaction of the NEC for a given $M$ requires the stronger
inequality \eqref{eq:NEC_inequality}.  For a concrete analysis
one must solve \eqref{eq:NEC_inequality} together with the
static condition; the result is generally a finite interval of
$a$ (if any) that depends on the ratio $M/l_0$.\\

\textbf{Strong Energy Condition.}  The SEC requires both
$\sigma + \mathcal{P} \geq 0$ and $\sigma + 2\mathcal{P} \geq 0$.
For the T-duality wormhole, using the surface pressure,
\begin{equation}
	\sigma(a) + 2\mathcal{P}(a) =
	\frac{M a (a^2 - 2l_0^2)}
	{2\pi (a^2 + l_0^2)^{5/2}\,
		\sqrt{1 - \dfrac{2M a^2}{(a^2 + l_0^2)^{3/2}}}} .
	\label{sigma+2P}
\end{equation}
The denominator is positive whenever the static condition holds,
so the sign of $\sigma+2\mathcal{P}$ is the sign of
$(a^2-2l_0^2)$.  Hence $\sigma+2\mathcal{P} \ge 0$ precisely
when $a \ge \sqrt{2}\,l_0$.  The first SEC condition
($\sigma+\mathcal{P}\ge0$) is the NEC discussed above.
Therefore, for a given $M$, the SEC holds for throat radii that
simultaneously satisfy \eqref{eq:NEC_inequality} and
$a \ge \sqrt{2}l_0$.  In particular, if the wormhole parameters
are such that the NEC is satisfied for $a$ larger than
$\sqrt{2}l_0$, then the SEC is also satisfied in that region.
This behaviour contrasts sharply with the classical
Schwarzschild thin-shell wormhole, where the SEC is always
violated~\cite{Poisson:1995sv}.  The T-duality regularisation thus
allows the shell to satisfy the strong energy condition for a
sufficiently large throat radius, provided the mass $M$ is
chosen appropriately.\\

\textbf{Weak and Dominant Energy Conditions.}  The WEC requires
$\sigma \geq 0$ in addition to the NEC.  Since $\sigma(a) < 0$
for all $a$ [cf.\ Eq.~\eqref{gen-surfenergyBH}], the WEC is
\emph{always} violated at the throat.  The DEC requires
$|\mathcal{P}| \leq |\sigma|$, which is also violated for a wide
range of parameters; however, a detailed analysis shows that the
DEC may become marginally satisfied for very large $a$, though
this does not affect the main conclusions.\\

Table~\ref{tab:energy_conditions} summarises the status of the
various energy conditions, assuming that the wormhole parameters
are chosen so that the NEC holds for $a > \sqrt{2}l_0$ (which is
possible, for instance, by taking $M$ in the appropriate range).
The only energy conditions that can be satisfied for a
sufficiently large throat radius are the NEC and the SEC, a
non-trivial consequence of the T-duality regularisation that
distinguishes this model from the classical Schwarzschild
thin-shell wormhole, for which both NEC and SEC are always
violated~\cite{Poisson:1995sv}.  Physically, the T-duality
regularisation softens the gravitational potential at small
radii, reducing the magnitude of the negative surface pressure
required to support the throat and allowing the shell to meet
the stronger energy constraints at large radii.

\begin{table}[h]
	\centering
	\caption{Energy conditions for the static T-duality wormhole,
		assuming parameters are such that the NEC (and hence SEC)
		holds for $a > \sqrt{2}\,l_0$.  The symbol \checkmark{}
		indicates satisfaction, $\times$ indicates violation.}
	\label{tab:energy_conditions}
	\begin{tabular}{c c c}
		\hline
		Energy condition & $a < \sqrt{2}\,l_0$ & $a > \sqrt{2}\,l_0$ \\
		\hline
		NEC ($\sigma + \mathcal{P} \geq 0$) & $\times$ & \checkmark{} \\
		WEC ($\sigma \geq 0$, NEC)          & $\times$ & $\times$ \\
		SEC ($\sigma + \mathcal{P} \geq 0$,
		$\sigma + 2\mathcal{P} \geq 0$) & $\times$ & \checkmark{} \\
		DEC ($|\mathcal{P}| \leq |\sigma|$) & $\times$ & $\times$ \\
		\hline
	\end{tabular}
\end{table}

\section{Equation of motion and linearized stability}
\label{secIV}

\subsection{Equation of motion of the throat}

From Eq.~\eqref{eq:sigma_dynamic}, the junction condition reads
\begin{equation}
	\sqrt{1 - \frac{b(a)}{a} + \dot{a}^2} = -2\pi a\,\sigma(a),
	\label{eq:junction_sqrt}
\end{equation}
where the negative sign follows from $\sigma(a) < 0$.  Squaring
and rearranging yields the equation of motion
\begin{equation}
	\dot{a}^2 + V(a) = 0,
	\label{eq:eom_potential}
\end{equation}
with the effective potential
\begin{equation}
	V(a) = 1 - \frac{b(a)}{a}
	- \left( \frac{m_s(a)}{2a} \right)^2,
	\label{defpotential}
\end{equation}
where $m_s(a) \equiv 4\pi a^2 \sigma(a)$ is the surface mass of
the thin shell.

Equation~\eqref{eq:eom_potential} describes the one-dimensional
motion of a point particle of unit mass and zero total energy in
the potential $V(a)$, with $a(\tau)$ the position, $\tau$ the
Newtonian time, and $\dot{a}$ the velocity.  The acceleration
follows as $\ddot{a} = -V'(a)/2$.

A static solution at $a = a_0$ requires
$V(a_0) = V'(a_0) = 0$.  The first condition gives the surface
mass at equilibrium:
\begin{equation}
	m_s(a_0) = -2a_0 \sqrt{1 - \frac{b(a_0)}{a_0}},
	\label{eq:ms_static_negative}
\end{equation}
where the negative sign is selected by consistency with
Eq.~\eqref{eq:sigma_dynamic}.  The second condition,
$V'(a_0) = 0$, yields the equilibrium relation
\begin{equation}
	m_s'(a_0) = \frac{m_s(a_0)}{2a_0}
	- \frac{1 - b'(a_0)}{\sqrt{1 - \dfrac{b(a_0)}{a_0}}} .
	\label{eq:ms_prime_equilibrium}
\end{equation}

Equation~\eqref{eq:ms_prime_equilibrium} is a purely geometric
consistency condition, independent of the specific equation of
state of the exotic matter.  As we shall see, the value of
$m_s'(a_0)$ determines the stability of the static configuration.
For completeness, the effective potential can be inverted to
give $m_s(a)$ in terms of $V(a)$,
\begin{equation}
	m_s(a) = -2a \sqrt{1 - \frac{b(a)}{a} - V(a)},
	\label{eq:ms_from_V}
\end{equation}
a formulation that is useful for constructing wormholes with
prescribed stability properties~\cite{Garcia:2011aa,MartinMoruno:2011rm,Bouhmadi-Lopez:2014gza}.

\subsection{Linearized stability analysis}
\label{sec:linearized}

We now carry out a detailed linearized stability analysis of the
static thin-shell wormhole configurations, following the unified
formalism developed by Garcia, Lobo, Martin-Moruno and
Visser (GLMV)~\cite{Garcia:2011aa,MartinMoruno:2011rm} (see also
Refs.~\cite{MartinMoruno:2011rm,Poisson:1995sv,Eiroa:2003wp,Lobo:2003xd,Musgrave:1995ka,Lobo:2025tph,Ishak:2001az}
for related applications).  The central idea of this approach is
to express the stability condition entirely in terms of the
normalised surface mass $\mu(a) = m_s(a)/a$ and its derivatives,
thereby achieving a clean separation between the geometric
contribution---fixed by the background metric---and the matter
contribution---governed by the equation of state of the exotic
fluid supporting the throat.  This formulation avoids the
redundancy of deriving the same stability inequality in multiple
equivalent forms and makes the geometric nature of the stability
threshold manifest.

\subsubsection{Taylor expansion and the stability criterion}

The dynamics of the throat are governed by the effective
potential~\eqref{defpotential}, which we rewrite in terms of the
normalised surface mass $\mu(a) = m_s(a)/a$ as
\begin{equation}
	V(a) = 1 - \frac{b(a)}{a} - \left( \frac{\mu(a)}{2} \right)^2.
	\label{eq:V_mu}
\end{equation}

For a static solution at $a = a_0$, the conditions
$V(a_0) = 0$ and $V'(a_0) = 0$ hold.  Expanding
$V(a)$ about $a_0$ and using the static conditions to eliminate
the zeroth- and first-order terms, we obtain
\begin{equation}
	V(a) = \frac{1}{2} V''(a_0)(a - a_0)^2
	+ \mathcal{O}\!\left[ (a - a_0)^3 \right],
	\label{eq:potential_quadratic}
\end{equation}
whereupon the equation of motion~\eqref{eq:eom_potential}
reduces to
\begin{equation}
	\dot{a}^2 = -\frac{1}{2} V''(a_0)(a - a_0)^2
	+ \mathcal{O}\!\left[ (a - a_0)^3 \right].
	\label{eq:eom_linearized}
\end{equation}
If $V''(a_0) < 0$, the potential has a local maximum and the
configuration is \emph{unstable}---any small perturbation causes
the throat radius to diverge exponentially from equilibrium at
the Lyapunov rate $\lambda = \sqrt{|V''(a_0)|/2}$.  Conversely,
if $V''(a_0) > 0$, the potential has a local minimum and the
configuration is \emph{stable}, with small perturbations
oscillating harmonically about $a_0$ at the characteristic
frequency $\omega_0 = \sqrt{V''(a_0)/2}$.  The marginal case
$V''(a_0) = 0$ corresponds to a bifurcation point and requires a
higher-order (nonlinear) stability analysis to determine the
fate of finite-amplitude perturbations.

\subsubsection{Static conditions in the $\mu$-formulation}

From Eq.~\eqref{eq:V_mu}, the static conditions
$V(a_0) = 0$ and $V'(a_0) = 0$ yield the equilibrium values of
the normalised surface mass and its first derivative:
\begin{align}
	\mu(a_0) &= -2 \sqrt{1 - \frac{b(a_0)}{a_0}}, \label{eq:mu_static} \\[4pt]
	\mu'(a_0) &= \frac{\left( \dfrac{b(a)}{a} \right)'_{a=a_0}}
	{\sqrt{1 - \dfrac{b(a_0)}{a_0}}}. \label{eq:mu_prime_static}
\end{align}

Equation~\eqref{eq:mu_static} determines the surface mass at the
throat; the negative sign reflects the fact that the shell
matter violates the weak energy condition.
Equation~\eqref{eq:mu_prime_static} is a purely geometric
consistency condition: it links the gradient of the surface mass
to the gradient of the metric function, independent of any
specific equation of state.  Any physically acceptable surface
mass function must satisfy this relation at the equilibrium
radius.

\subsubsection{Second derivative and the GLMV master stability condition}

The stability of the static configuration is determined by the
sign of $V''(a_0)$.  Differentiating Eq.~\eqref{eq:V_mu} twice
and evaluating at $a_0$ using the static
conditions~\eqref{eq:mu_static}--\eqref{eq:mu_prime_static}, we
obtain the central result of the linearized analysis:
\begin{eqnarray}
	V''(a_0) &=& \sqrt{1 - \frac{b(a_0)}{a_0}} \; \mu''(a_0)
	- \frac{1}{2}\,
	\frac{\left[ \left( \dfrac{b(a)}{a} \right)'_{a=a_0} \right]^2}
	{\left( 1 - \dfrac{b(a_0)}{a_0} \right)}
	\nonumber \\
	&& \qquad \qquad - \left( \dfrac{b(a)}{a} \right)''_{a=a_0}.
	\label{eq:V_double_prime_mu_corrected}
\end{eqnarray}

This expression is remarkably transparent.  The first term
involves the second derivative of the normalised surface mass,
weighted by the positive factor $\sqrt{1-b(a_0)/a_0}$.  The
remaining terms involve only the combination $b(a)/a$ and its
derivatives; they encode the contribution of the background
spacetime geometry to the stability of the throat.  Because
these geometric terms are negative (they subtract from the
$\mu''$ term), the geometry tends to destabilise the wormhole
unless the matter contribution $\mu''(a_0)$ is sufficiently
large and positive.

Setting $V''(a_0) > 0$ and rearranging yields the master
stability condition in the form derived by Garcia, Lobo,
Martin-Moruno and Visser~\cite{Garcia:2011aa,MartinMoruno:2011rm}:
\begin{equation}
	\mu''(a_0) \geq \mathcal{G}(a_0) =
	\frac{1}{2}\,
	\frac{\left[ \left( \dfrac{b(a)}{a} \right)'_{a=a_0} \right]^2}
	{\left( 1 - \dfrac{b(a_0)}{a_0} \right)^{3/2}}
	+ \frac{\left( \dfrac{b(a)}{a} \right)''_{a=a_0}}
	{\sqrt{1 - \dfrac{b(a_0)}{a_0}}} .
	\label{eq:master_stability_GLMV_corrected}
\end{equation}

The right-hand side is the \emph{geometric stability threshold},
\begin{equation}
	\mathcal{G}(a) \equiv
	\frac{1}{2}\,
	\frac{\left[ \left( \dfrac{b(a)}{a} \right)' \right]^2}
	{\left( 1 - \dfrac{b(a)}{a} \right)^{3/2}}
	+ \frac{\left( \dfrac{b(a)}{a} \right)''}
	{\sqrt{1 - \dfrac{b(a)}{a}}},
	\label{eq:geometric_threshold_GLMV_corrected}
\end{equation}
which depends exclusively on the background spacetime geometry
through the single dimensionless combination $b(a)/a$ and its
first two derivatives.  The master
condition~\eqref{eq:master_stability_GLMV_corrected} is the
compact, necessary, and sufficient criterion for linearized
stability.  Regions where $\mathcal{G}(a) < 0$ are geometrically
favourable: the inequality is then automatically satisfied for
any $\mu''(a_0) \geq 0$, placing no restriction on the equation
of state beyond the requirement that the surface mass be a
convex function of the throat radius.  Large positive values of
$\mathcal{G}(a)$, in contrast, demand a correspondingly large
$\mu''(a_0)$, which translates into stringent constraints on the
shell matter.

Several features of the GLMV formulation deserve emphasis.
First, it makes the geometric nature of the stability threshold
completely manifest: $\mathcal{G}(a)$ is constructed entirely
from $b(a)/a$, with no reference to surface stresses or
equation-of-state parameters.  Second, it unifies the various
equivalent forms of the stability condition that appear in the
literature~\cite{Poisson:1995sv,Eiroa:2003wp,Lobo:2003xd,Musgrave:1995ka,Lobo:2025tph,Ishak:2001az} into a single,
compact inequality.  Third, the normalised mass $\mu(a)$ is
dimensionless, simplifying the dimensional analysis and
facilitating comparison across different regularised metrics.

\section{Master stability conditions and application to the T-duality wormhole}\label{secV}

\subsection{Geometric threshold for the T-duality wormhole}

For the T-duality metric~\eqref{def:b},
$b(a) = 2M a^3/(a^2 + l_0^2)^{3/2}$, the relevant
combinations are
\begin{align}
	\frac{b(a)}{a} &= \frac{2M a^2}{(a^2 + l_0^2)^{3/2}}, \\[4pt]
	\left( \frac{b(a)}{a} \right)' &= \frac{2M a (2l_0^2 - a^2)}{(a^2 + l_0^2)^{5/2}}, \\[4pt]
	\left( \frac{b(a)}{a} \right)'' &= \frac{2M (2l_0^4 - 11l_0^2 a^2 + 2a^4)}{(a^2 + l_0^2)^{7/2}}.
\end{align}

Substituting these into Eq.~\eqref{eq:geometric_threshold_GLMV_corrected}
yields the explicit geometric stability threshold for the
T-duality wormhole:
\begin{align}
	\mathcal{G}(a) &=
	\frac{2M^2 a^2 (2l_0^2 - a^2)^2}
	{(a^2 + l_0^2)^5\,
		\left( 1 - \dfrac{2M a^2}{(a^2 + l_0^2)^{3/2}} \right)^{3/2}}
	\nonumber \\
	&\quad + \frac{2M (2l_0^4 - 11l_0^2 a^2 + 2a^4)}
	{(a^2 + l_0^2)^{7/2}\,
		\sqrt{1 - \dfrac{2M a^2}{(a^2 + l_0^2)^{3/2}}}} .
	\label{eq:geometric_threshold_T_GLV}
\end{align}

This expression is algebraically equivalent to the threshold
derived in the traditional Poisson--Visser formalism \cite{Poisson:1995sv} (see
Eq.~\eqref{eq:geometric_threshold_GLMV_corrected}), but its structure is
more compact and its geometric origin is more transparent.  It
is valid for all throat radii satisfying the static
condition~\eqref{conditionMshell},
$M < (a^2 + l_0^2)^{3/2}/(2a^2)$, which ensures that the square
roots in the denominators are real and positive.


The geometric stability threshold $\mathcal{G}(a)$ given by
Eq.~\eqref{eq:geometric_threshold_T_GLV} can be expressed in a
manifestly dimensionless form by introducing the variables
\begin{equation}
	x \equiv \frac{2M}{a}, \qquad y \equiv \frac{l_0}{a},
	\label{eq:dimensionless_variables}
\end{equation}
and defining $F(x,y) \equiv a^2\,\mathcal{G}(a)$.  

The physical meaning of these variables is as follows:
(i) $x = 2M/a$ is the ratio of the Schwarzschild radius of the
central mass (had it been a black hole) to the throat radius.
In the classical limit $l_0 \to 0$, the condition $x<1$
corresponds to having the throat outside the would‑be horizon.
(ii) $y = l_0/a$ quantifies the relative importance of the
T‑duality regularisation scale. When $y \ll 1$, the throat is
much larger than the quantum smearing length; the geometry
approaches the classical Schwarzschild case. When $y \sim 1$,
quantum effects become significant.
(iii) The product $a^2 \mathcal{G}(a)$ is dimensionless because
$\mathcal{G}(a)$ has dimensions of inverse length squared
(since $V''(a)$ does). Hence $F(x,y)$ is a pure number
depending only on the two ratios $x$ and $y$.

Taking into account the static condition
$M < (a^2 + l_0^2)^{3/2}/(2a^2)$, the domain of the variables is
$0 < x < (1+y^2)^{3/2}$ and $0<y<\infty$.  The upper bound on
$x$ ensures that the throat remains outside any horizon, so the
wormhole is traversable.

The stability regions are determined by the following master
equation:
\begin{equation}
	F(x,y)= a_0^2 \, \mu''(a_0) \geq a_0^2 \, \mathcal{G}(a_0) .
	\label{eq:master_stability_GLV2}
\end{equation}
Physically, the left‑hand side $a_0^2 \mu''(a_0)$ encodes the
response of the exotic matter to a change in throat radius.
Using the conservation equation~\eqref{eq:sigma_prime_conservation}
and the definition $\mu(a)=m_s(a)/a$, one can show that
\begin{equation}
	\mu''(a_0) = \frac{8\pi(1+2\eta)(\sigma+\mathcal{P})}{a_0},
\end{equation}
where $\eta = \mathcal{P}'/\sigma'$ is the squared sound speed
of the shell material.  Thus $a_0^2\mu''(a_0)$ is directly
related to the equation of state through $\eta$ and the
combination $\sigma+\mathcal{P}$.  The inequality therefore
states that the matter must be “stiff enough” (large positive
$\mu''$) to overcome the destabilising tendency of the
background geometry, quantified by $\mathcal{G}(a_0)$.  When
$\mathcal{G}(a_0)$ is negative, the geometry itself already
favours stability, and any convex surface mass function
($\mu''\ge0$) suffices.  When $\mathcal{G}(a_0)$ is positive,
the matter must satisfy a positive lower bound on $\mu''$,
which translates into a constraint on the equation of state.

Substituting $M = x a/2$ and $l_0 = y a$ into
Eq.~\eqref{eq:geometric_threshold_T_GLV} yields the explicit
dimensionless geometric threshold:
\begin{align}
	F(x,y) &=
	\frac{x^2 (2y^2 - 1)^2}
	{2\,(1 + y^2)^{11/4}\,
		\bigl[ (1 + y^2)^{3/2} - x \bigr]^{3/2}}
	\nonumber \\
	&\quad + \frac{x\,(2y^4 - 11y^2 + 2)}
	{(1 + y^2)^{11/4}\,
		\bigl[ (1 + y^2)^{3/2} - x \bigr]^{1/2}} .
	\label{eq:F_direct}
\end{align}
This expression is valid for all $(x,y)$ in the domain
$0 < x < (1+y^2)^{3/2}$, $y>0$.

\begin{figure}[t!]
	\centering
	\includegraphics[scale=0.3]{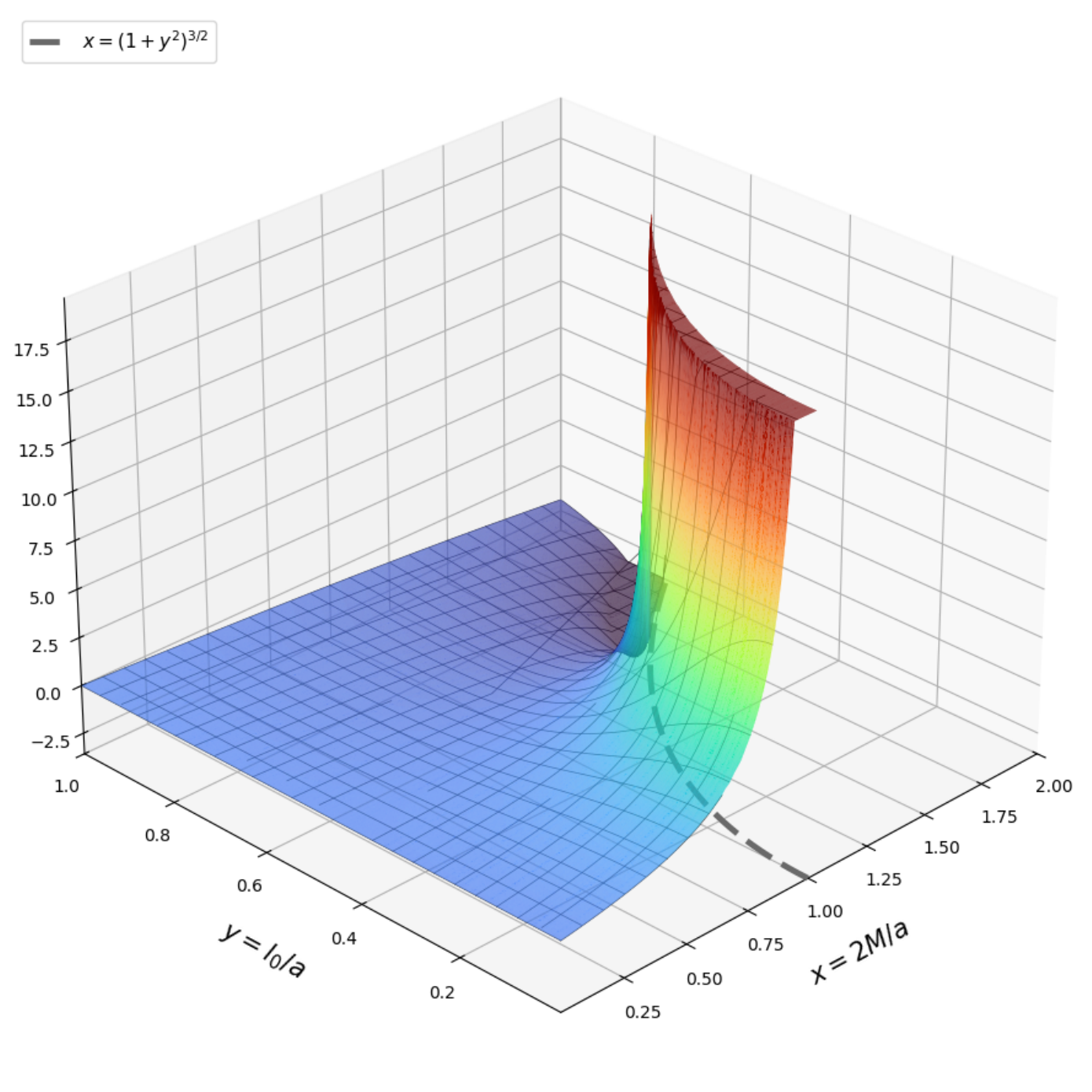}
	\caption{The solid contour surface depicts the geometric stability threshold $F(x,y)$ as a function of the dimensionless variables $x = 2M/a$ and $y = l_0/a$. Stability regions lie above this surface. In regions where $F(x,y)$ is negative, the geometry is stabilising, and any convex surface mass function ($\mu''\ge0$) yields stability.}
	\label{fig:stability}
\end{figure}

Eqs.~\eqref{eq:master_stability_GLV2} and \eqref{eq:F_direct}
are the master equations that must be analysed to determine the
stability of the T‑duality wormhole.  Their graphical
representation (Fig.~\ref{fig:stability}) shows the region in
the $(x,y)$ plane where the geometric threshold is below a given
value of $a_0^2\mu''(a_0)$.  For a physically plausible
equation of state (e.g., a barotropic fluid with a fixed sound
speed), one can overlay the corresponding $a_0^2\mu''(a_0)$
contour and identify stable configurations.  The plot highlights
that the T‑duality regularisation opens up islands of stability
that are absent in the classical Schwarzschild case, where
$x<1$ and $y=0$ leads to a strictly positive $\mathcal{G}(a)$
for all $a$.

\subsection{Qualitative behaviour of the geometric threshold}
\label{sec:qualitative_G}

The stability condition derived from the GLMV formalism is
\(\mu''(a_0) \ge \mathcal{G}(a_0)\), where \(\mu(a)=m_s(a)/a\) is the
normalised surface mass and \(\mathcal{G}(a)\) is the geometric
threshold given by Eq.~\eqref{eq:geometric_threshold_GLMV_corrected}.
Because \(\mu(a_0)<0\) for the exotic shell, a positive
\(\mu''(a_0)\) tends to stabilise the configuration, while a
negative \(\mu''(a_0)\) destabilises it.  The sign and magnitude
of \(\mathcal{G}(a_0)\) are therefore crucial: if
\(\mathcal{G}(a_0) < 0\), the inequality is automatically
satisfied for any \(\mu''(a_0) \ge 0\) (the typical case for
causal matter), meaning the wormhole is \emph{unconditionally
	stable} against radial perturbations.  Conversely, if
\(\mathcal{G}(a_0) > 0\), a sufficiently large positive
\(\mu''(a_0)\) is required, which translates into a stiff
equation of state for the shell.  In the limit
\(\mathcal{G}(a_0) \to +\infty\) the configuration is always
unstable because no finite \(\mu''(a_0)\) can satisfy the
inequality.

We now analyse the behaviour of \(\mathcal{G}(a)\) for the
T‑duality wormhole, using the explicit expression
\eqref{eq:geometric_threshold_T_GLV}.  The analysis is carried
out for a representative mass scale, e.g. \(M/l_0 = 1\), but the
qualitative features are generic.

\medskip\noindent
\textbf{Definition of the minimum throat radius.}
The static wormhole exists only for throat radii \(a\) such that
\(f(a)=1-b(a)/a > 0\).  The inequality
\begin{equation}
	M < \frac{(a^2 + l_0^2)^{3/2}}{2a^2}
\end{equation}
defines the allowed region.  The minimum allowed radius
\(a_{\rm min}\) is the largest positive root of the equality
\(M = (a^2+l_0^2)^{3/2}/(2a^2)\), which corresponds to
\(f(a_{\rm min}) = 0\).  For a given \(M\) and \(l_0\),
\(a_{\rm min}\) is the outer horizon of the regularised black
hole.  As the throat approaches this radius from above, the
redshift factor \(f(a)\) tends to zero.

\medskip\noindent
\textbf{Small throat radii (\(a_0\) close to \(a_{\rm min}\)).}
As \(a_0 \to a_{\rm min}^+\), the factor
\(1 - b(a_0)/a_0 = f(a_0) \to 0^+\).  The geometric threshold
\(\mathcal{G}(a_0)\) contains terms proportional to
\(f(a_0)^{-3/2}\) and \(f(a_0)^{-1/2}\) (see
Eq.~\eqref{eq:geometric_threshold_GLMV_corrected}), so it diverges to
\(+\infty\).  Consequently, the stability condition
\(\mu''(a_0) \ge \mathcal{G}(a_0)\) cannot be satisfied for any
finite \(\mu''(a_0)\).  The throat is therefore \emph{unstable}
when it is located too close to the would‑be horizon.  This
behaviour is universal for thin‑shell wormholes: the
gravitational attraction becomes arbitrarily strong as the
throat approaches the horizon radius, and no finite stiffness
can prevent collapse or runaway expansion.

\medskip\noindent
\textbf{Intermediate throat radii (\(a_0 \sim l_0\)).}
As \(a_0\) increases from \(a_{\rm min}\), the geometric threshold
decreases from \(+\infty\).  For certain values of \(M/l_0\),
\(\mathcal{G}(a_0)\) may become negative over a finite interval.
This happens when the numerator in the second term of
Eq.~\eqref{eq:geometric_threshold_T_GLV} (which contains the
factor \(2l_0^4-11l_0^2a^2+2a^4\)) changes sign, and the positive
first term is sufficiently small.  For the representative case
\(M/l_0 = 1\), a numerical evaluation of
Eq.~\eqref{eq:geometric_threshold_T_GLV} shows that
\(\mathcal{G}(a_0) < 0\) for a range of
intermediate radii (e.g., approximately \(1.5\,l_0 \lesssim a_0
\lesssim 3\,l_0\)).  In this window the stability condition
\(\mu''(a_0) \ge \mathcal{G}(a_0)\) is automatically satisfied
for any \(\mu''(a_0) \ge 0\) (i.e., for any physically
reasonable shell material), leading to \emph{unconditional
	stability}.  This behaviour is a direct consequence of the
T‑duality regularisation; it is absent in the Schwarzschild
wormhole, where \(\mathcal{G}(a_0)\) is always positive.

\medskip\noindent
\textbf{Large throat radii (\(a_0 \gg l_0\)).}
In the asymptotic regime \(a_0 \gg l_0\), the T‑duality metric
approaches the Schwarzschild form: \(b(a_0)/a_0 \simeq 2M/a_0\).
The geometric threshold then behaves as
\begin{equation}
	\mathcal{G}(a_0) \simeq \frac{4M}{a_0^3}
	+ \mathcal{O}\!\left(\frac{M^2}{a_0^4}\right) > 0 .
\end{equation}
This positive asymptotic value implies that for very large
throats, stability requires \(\mu''(a_0)\) to be sufficiently
large.  Because \(\mu(a_0) \approx -2a_0\sqrt{1-2M/a_0}\), a
positive \(\mu''(a_0)\) corresponds to a stiff equation of state
(high sound speed).  Note that the same asymptotic behaviour
holds for the Schwarzschild wormhole; the leading term is
\(4M/a^3 > 0\).  The T‑duality regularisation does not alter
this leading‑order positive term, but it can suppress
\(\mathcal{G}(a)\) at intermediate scales, creating the
unconditional stability window mentioned above.

\medskip\noindent
\textbf{Summary of the qualitative behaviour.}
The geometric threshold \(\mathcal{G}(a)\) for the T‑duality
wormhole exhibits three distinct regimes:
\begin{itemize}
	\item A near‑horizon regime (\(a_0 \to a_{\rm min}^+\)) where
	\(\mathcal{G}\to+\infty\) and the wormhole is always unstable.
	\item An intermediate regime (\(a_0\) of order \(l_0\)) where
	\(\mathcal{G}(a_0)\) can become negative, yielding unconditional
	stability for any \(\mu''(a_0)\ge 0\).
	\item A large‑radius regime (\(a_0 \gg l_0\)) where
	\(\mathcal{G}(a_0) \sim 4M/a_0^3 > 0\), so stability requires a
	sufficiently stiff shell.
\end{itemize}
These features are illustrated in Fig.~\ref{fig:stability}. The existence of a window with
\(\mathcal{G}<0\) is a unique signature of the T‑duality
regularisation and distinguishes this model from the classical
Schwarzschild thin‑shell wormhole, for which \(\mathcal{G}\) is
everywhere positive.

\subsection{The role of the regularisation scale $l_0$}
\label{sec:role_l0}

The T-duality regularisation introduces the length scale
$l_0$, which fundamentally alters the stability properties
relative to the classical Schwarzschild thin-shell wormhole.
For the Schwarzschild case ($l_0 = 0$), the geometric
threshold $\mathcal{G}_{\rm Schw}(a_0)$ is positive for all
$a_0 > 2M$.  Since the stability condition requires $\mu''(a_0) \ge \mathcal{G}(a_0)$ and $\mu''(a_0)$ is typically non‑negative for physical matter, a positive $\mathcal{G}(a_0)$ imposes a non‑trivial lower bound on the convexity of the surface mass.  No region of unconditional stability (where $\mathcal{G}(a_0)<0$) exists in the Schwarzschild case.

The T-duality regularisation modifies the short-distance
behaviour of the metric through the replacement
$b(a) \sim 2M a^3/l_0^3$ for $a \ll l_0$,
so that $b(a)/a \sim 2M a^2/l_0^3 \to 0$ as
$a \to 0$.  The metric therefore describes a regular core
rather than a singularity.  This modification allows the
geometric threshold to become negative over a range of
intermediate radii---a feature that must be verified
numerically from
Eq.~\eqref{eq:geometric_threshold_T_GLV}---potentially
creating a window of unconditional stability (i.e., $\mathcal{G}(a_0)<0$) that is absent
in the classical case.  At large radii ($a \gg l_0$), the
metric asymptotes to the Schwarzschild form
$b(a)/a \sim 2M/a$, and the geometric threshold behaves
as $\mathcal{G}(a) \sim 4M/a^3 > 0$, so that stability requires a sufficiently large $\mu''(a_0)$, in qualitative
agreement with the Schwarzschild case.

Physically, the regularisation scale acts as an effective
cutoff on the gravitational field: at distances
$r \lesssim l_0$, the classical $1/r$ divergence of the
Schwarzschild potential is replaced by a finite, regular
core.  This reduces the magnitude of the negative surface
pressure required to support the throat and, consequently,
may lessen the severity of the stability constraints in
the intermediate regime.  At large distances, the
T-duality wormhole behaves similarly to its Schwarzschild
counterpart, with stability requiring a sufficiently convex surface mass function.

These results demonstrate that the T-duality wormhole
possesses a rich stability structure: a window of instability
near the minimum throat radius, where $\mathcal{G}(a_0) \to +\infty$;
a possible window of unconditional stability at intermediate
radii (depending on the value of $M/l_0$), where $\mathcal{G}(a_0) < 0$;
and a region of conditional stability at large radii, where
$\mathcal{G}(a_0) > 0$ and a sufficiently large $\mu''(a_0)$ is required.

\section{Discussion and Conclusion}\label{sec:conclusion}

In this work we have constructed a traversable thin‑shell wormhole
by joining two T‑duality regularised spacetimes.  The quantum‑corrected
metric replaces the classical Schwarzschild singularity with a regular
core, allowing a well‑defined cut‑and‑paste construction.  Using the
Israel junction conditions we derived the surface stress‑energy tensor
of the shell and obtained the equation of motion for the throat.
A linearised stability analysis was performed within the unified GLMV
formalism, which expresses the stability condition solely in terms of
the geometric threshold $\mathcal{G}(a)$ and the second derivative of
the normalised surface mass $\mu(a)$.

The surface energy density of the shell is negative for all static
configurations, confirming that the wormhole throat requires exotic
matter that violates the weak energy condition.  Nevertheless, the
T‑duality regularisation allows the null and strong energy conditions
to be satisfied for sufficiently large throat radii, provided the mass
parameter is chosen appropriately.  This behaviour contrasts sharply
with the classical Schwarzschild thin‑shell wormhole, where the NEC
and SEC are always violated.

The stability analysis reveals that the dynamics of the throat reduce
to a one‑dimensional effective potential.  The condition for linearised
stability is $V''(a_0)>0$, which translates into the master inequality
$\mu''(a_0)\ge\mathcal{G}(a_0)$.  The geometric threshold $\mathcal{G}(a)$
depends only on the background metric.  For the Schwarzschild case
($l_0=0$) it is positive for all throat radii, so stability requires
a sufficiently large $\mu''(a_0)$ (a stiff equation of state) and no
unconditional stability window exists.  The T‑duality regularisation
introduces the length scale $l_0$, which softens the gravitational
potential at short distances: for $a\ll l_0$ the metric function
$b(a)/a$ tends to zero, describing a regular core.  As a consequence,
the geometric threshold can become negative over a range of intermediate
radii (e.g., for $M/l_0=1$, roughly $1.5l_0\lesssim a_0\lesssim 3l_0$).
In this window $\mathcal{G}(a_0)<0$, so the stability condition is
automatically satisfied for any $\mu''(a_0)\ge0$; the wormhole is
unconditionally stable against radial perturbations.  At large radii
($a_0\gg l_0$) the metric asymptotes to the Schwarzschild form, and
$\mathcal{G}(a_0)$ becomes positive again, requiring a sufficiently
convex surface mass function.  Thus the T‑duality wormhole exhibits
three regimes: a near‑horizon unstable region, an intermediate
unconditional stability window (for suitable $M/l_0$), and a
large‑radius conditional stability region.

The role of the regularisation scale $l_0$ is therefore twofold:
it cures the curvature singularity and, more importantly, it enriches
the stability landscape by creating a window where the geometry
itself favours stability.  This unique signature distinguishes the
T‑duality wormhole from its classical counterpart and suggests that
quantum‑gravity motivated modifications can render traversable
wormholes more plausible from a stability perspective.

Several directions for future research naturally emerge.  The
present analysis focused on radial perturbations; extending the
study to non‑spherical deformations would reveal whether the
unconditional stability window survives against higher‑multipole
modes.  Other regularisation schemes (e.g., non‑commutative geometry
or asymptotically safe gravity) could be compared to identify
generic features of regular wormholes.  Explicit, physically
motivated equations of state for the exotic shell matter should be
constructed to check whether they satisfy the stability inequality.
Observational signatures such as gravitational lensing, shadows, or
echoes in gravitational wave signals could be computed for the
T‑duality wormhole and contrasted with black hole predictions,
providing concrete tests for future detectors.  Finally, asymmetric
junctions (different masses or regularisation scales on each side)
would introduce an additional parameter that might enlarge or shrink
the stability window and could give rise to a net gravitational force
on the shell.  These investigations will help assess whether
regularised thin‑shell wormholes can serve as viable astrophysical
alternatives to black holes.

\acknowledgments

FSNL acknowledges support from the Funda\c{c}\~{a}o para a Ci\^{e}ncia e a Tecnologia (FCT) Scientific Employment Stimulus contract with reference CEECINST/00032/2018, and funding through the research grant UID/04434/2025.
MER thanks Conselho Nacional de Desenvolvimento Cient\'ifico e Tecnol\'ogico - CNPq, Brazil, for partial financial support.



\end{document}